# GPSPiChain-Blockchain based Self-Contained Family Security System in Smart Home

Ali Raza[1][0000-0002-5303-7005] and Lachlan Hardy[1] [0000-0002-3926-6418]

Erin Roehrer[1] [0000-0001-6211-7747] and Soonja Yeom[1] [0000-0002-5843-101X]

Byeong Ho Kang[1] [0000-0003-3476-8838]

[1] School of ICT, Syndicate of TED, University of Tasmania, Sandy Bay, Australia
Ali.raza@utas.edu.au

**Abstract.** With advancements in technology, personal computing devices are better adapted for and further integrated into people's lives and homes. The integration of technology into society also results in an increasing desire to control who and what has access to sensitive information, especially for vulnerable people including children and the elderly. With blockchain coming in to the picture as a technology that can revolutionize the world, it is now possible to have an immutable audit trail of locational data over time. By controlling the process through inexpensive equipment in the home, it is possible to control whom has access to such personal data. This paper presents a blockchain based family security system for tracking the location of consenting family members smart phones. The locations of the family members smart phones are logged and stored in a private blockchain which can be accessed through a node installed in the family home on a computer. The data for the whereabouts of family members stays within the family unit and does not go to any third party. The system is implemented in a small scale (one miner and two other nodes) and the technical feasibility is discussed along with the limitations of the system. Further research will cover the integration of the system into a smart home environment, and ethical implementations of tracking, especially of vulnerable people, using the immutability of blockchain.

**Keywords:** Blockchain, Security, Global Positioning System (GPS).

## 1 Introduction

As people acquire more technology, the role of the home is highlighted as a place of safety, a place of storage, and the hub of interaction for a family. This is achieved through the increasing use of technology to provide security and safety for the home and the people inside it. The desire for security is something all animals possess [1]. Those that tend to make static locations like dens, burrows, or nests may consider the location to be important as an aspect of security. For some humans, location has become a factor in increasing security, as can be seen with gated communities. But many humans are unable to live in such secure locations and have to rely on other means to provide that feeling of security. What provides the feeling of security in a technology rich environment? One answer is arguably information of where and when, and the knowledge that such information is verifiably true [2].

The quest for information that is collected and used to identify the whereabouts of a member of the family is an important one. As there are many issues that come from a family member being in the wrong location and having no active means of alerting family. Other technologies that have been investigated involve tracking through Radio-Frequency Identification (RFID) and Global Positioning System (GPS) which can work for animals or family members where tampering isn't a concern [3-5].

But one of the main problems with such technology for family protection is the concern of data tampering. Previous solutions to providing security and attempting to avoid data tampering have involved third-party vendors that host the tracking technology on their systems such as vehicle tracking [6, 7]. Such systems have problems from a security perspective, as the data can be acquired from the source server in a broad attack rather than a targeted attack against an individual's family. But one of the most recent problems with such systems comes from the value of data for artificial intelligence-based purposes, like machine learning, where issues related to protecting anonymity and privacy are increasing [8]. Current cloud-based digital platforms have difficulty in protecting an individual entities data through by separating it from other clients, while also retaining the validity and just access to the data [9].

As personal and family tracking becomes more prevalent, there will be an increasing demand to decrease the scale of involved systems to protect privacy, while also increasing the verifiability of the data to avoid easy tampering.

## 2 Background and Technologies

### 2.1 Family Security

There are two core aspects of family security which are safety and verifiability; safety means knowing that the family member is safe in their current situation [10]; and verifiability means knowing the information is accurate and meaningful [11]. Prior to the introduction of modern technology, which allows rapid social vetting and instant communication, safety through trust was difficult to establish [12]. As communication became faster with better technology, the trustworthiness of information proportionally increased. More recently, security was added to the role of safety and verifiability through the use of technology [13]. Technologies for tracking like GPS and RFID were slowly incorporated into personal security, which have reduced the scale of safety and verifiability down to the personal level [4].

Use of these technologies has had problems, as each innovation has been susceptible to increasingly sophisticated attempts to circumvent security and verifiability. Such attempts have ranged from the non-technical social engineering to more technical interception of signals or data manipulation [14]. The last factor, changing the data after transmission, has been of greater concern in the last two decades due to mutability concerns of the data relative to its volume [15].

The mutability factor ties in directly to the verifiability aspect of family security, if the record shows one thing then it should mean the record can be trusted, rather than having some doubts over its validity. This concern has existed for as long as personal home-based systems have existed, where the data logs can be changed by an enterprising party to remove verifiability [16]. The key challenge for preventing obfuscation of the data is knowing what to look for and how to identify its occurrence in the system. Often this was difficult as the one whom installed the system may not be the one most familiar with its operation, for example circumventing internet filtering is easier for those whom are technologically active than those whom are not [17]. One challenge of personal and family security in a home-based system is keeping the system operating while making the output immutable.

### 2.2 Blockchain.

Blockchain is a distributed ledger system consisting of a sequence of blocks which are appended in a peer-to-peer style of links, similar to the visual concept of a chain [18]. This means that the information entered in to a blockchain once, will stay there forever. Each block has a hash of its previous block thus making a chain linked through hashes. This technology was introduced by keeping the Byzantine problem in mind where nodes do not trust each other [19]. A block generally has a header and a body with content as is shown in Fig 1.

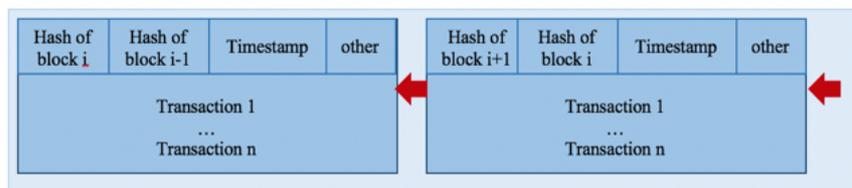

**Fig. 1.** Blockchain showing the blocks of data [20]

Each transaction is validated by each node of the network before adding it to the blockchain, ensuring a consensus is reached for every transaction. This resolves the single point of failure issues existing in traditional databases. In order to tamper, one would have to tamper with all the ledgers stored in at least 51% of the nodes in the network. Based on the permission style, blockchain has three different varieties which are public, private, and consortium. Public being permission-less, such that anyone can join and leave the network without permission. In contrast, private is a totally permission-based blockchain, and a node needs permission to be added in to the blockchain network. Consortium is a combination of public and private blockchains, such that some parts of the blockchain could be made public while some parts could be restricted to specific nodes [21].

Consensus algorithms are an important part of the blockchain since they bring trustworthiness to an untrusted network. There are different consensus algorithms, the most common are Proof of Work (PoW) [22], Proof of Stake (PoS) [23] and Practical Byzantine Fault Tolerant (pBFT) [19].

Different versions of blockchain technology have emerged since the proposition of Bitcoin in 2009. The original Bitcoin is considered to be Blockchain 1.0 and was solely based on cryptocurrency. People often get confused with the concept of Bitcoin, thinking of it as just a cryptocurrency rather than the underlying technology of bitcoin, which is a technology stack. The stack consists of a platform for transferring cryptocurrency, and a protocol for communication in that platform. The second version of blockchain emerged in the form of Ethereum, which introduced smart contracts.

A smart contract is a piece of code which executes automatically on pre-defined conditions becoming true. The third version of blockchain added decentralized applications (Dapps) along with smart contracts, allowing blockchain to be used in other domains but not limited to supply chain management, IoT, and finance. In the third version of blockchain systems, several other platforms came into existence using different approaches and features [24-26]. While blockchain is a distributed ledger system, it does not contain innate data that could be useful for a security-based system. It simply acts as the storage and security mechanism for such data.

## 2.3    Tracking

GPS as a tracking tool has been well documented over many decades, from large object tracking like boats down to the personal tracking of recent times [27]. The role of GPS is to receive the location data and pass it to an application to transmit it to another receiving device. Technology has been added to GPS to handle encryption, storage, and specialized purposes either as single additions or as part of multi-layered applications [28].

In adding GPS to these other technologies, new systems emerge to solve problems. Some previous problems come from the size of the devices, which was solved through shrinking the technology over time which resulted in reduced cost [29]. Other problems come from interfacing GPS with different protocols and devices to enable meaningful use of the data; as the boom in tracking using GPS has shown, this is no longer a problem.

However, with the decrease in costs and increase in desire to track location, another problem is emerging in the Internet of Things (IoT) space. As homes are becoming interconnected with technology resulting in smart-home systems, a natural extension of such technology is tracking people and property for security purposes outside of the home, using cheap but effective technology.

# 3    Problem Outline

Our proposed solution to the problem of tracking for personal security in a smart home environment using cheap resources ties into the use of a blockchain system based on a Raspberry Pi 3 connected to GSM SIM900 module, a laptop and a smart home miner (Server). While not all children and adults own such technology, the rate of ownership in developed nations is large [30] and growing at an increasing rate in developing nations [21]. As the proposed solution tracks the phone through the use of GPS, it is assumed that the phone will be in possession of the owner (family member). The solution will use the GPS data from the application contained on an android phone to trace the phone location and by proxy the person holding the phone.

The primary premise for this solution is based on the increasing desire for personal security of family members, and it further ties in to the security that comes from controlling the process in-house rather than through a third-party. As this process requires a family-based process to install the application on the phones, it is assumed that this will be used by consenting parties. As such the use of this process is outside the scope of this paper. The benefit of using a blockchain to encapsulate the process of security is the immutability of the data, which counteracts potential data change from within and without the home.

The immutable factor results in a tamper-less system that can capture the data from a small family unit and associated property while still having some scalability factor for the increasing demands of families and the emerging nature of IoT based smart homes. The immutability factor also deals with the current problem of verifiability of the data, as regardless of technical knowledge, there isn't currently a way of changing the data once it has been integrated into the chain as blockchain is append only.

# 4    Methodology design

## 4.1    Proposed Architecture

Our proposed architecture design is based on the outline shown in Fig 2. which involves four distinct items:

— Raspberry Pi 3
— GSM SIM900
— Miner (Server)
— Android phone (numbers depends on family members)

The transmitter side being the users/family members having android phones with an application installed in their phones which sends their GPS coordinates (longitude and latitude) of their current location after every 30 seconds via SMS to the receiver, which is a GSM SIM900 module connected to a Raspberry Pi 3 (RPi3). GSM SIM900 receives texts from all the family members/users containing their GPS Co-ordinates which are then passed to the RPi3.

A private blockchain was created for the smart-home with a RPi3 and a server (as the miner) as its nodes. The parent's laptop is also a node and controls this blockchain. The parents can add a new device and remove an existing device from blockchain. Unlike public blockchain, which is decentralize in terms of management, this private blockchain is centrally controlled by the parents in the family.

The RPi3 lacked the processing power to be a miner, as such a server was used as a miner in the private blockchain. After receiving data from the GSM SIM900, the RPi3 then sends the coordinates to the blockchain. Which after verifying the transaction containing coordinates of a phone by miner, adds them to the blockchain.

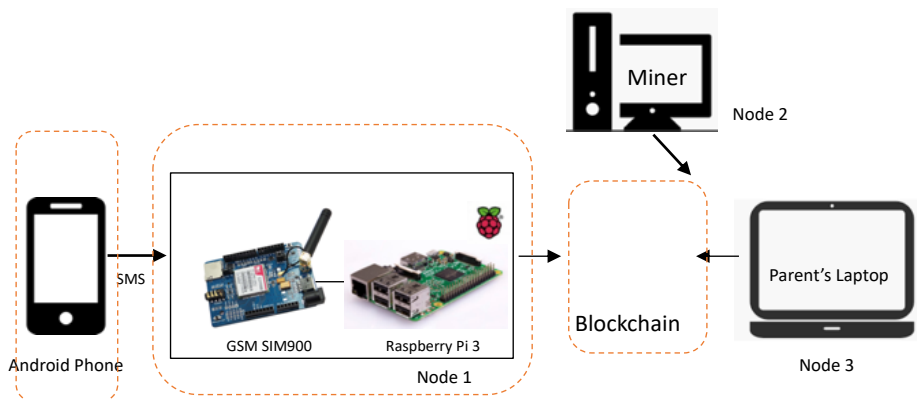

**Fig. 2.** Overall architecture of the system

The whole system can be subdivided into three different processes which will be outlined below.

### 4.2    Registration of a New Device

As shown in Fig 3 the parents laptop acts as the central control point for adding devices to the blockchain through the miner. The IMEI (International Mobile Equipment Identity) and phone number need to be manually added to the blockchain via miner which registers a device. The international standard for finding the IMEI on a phone is to dial *#06# which returns a 15-digit unique identifier. The phone number is registered to the sim card used by the phone through a telecommunications company.

The purpose of the smart contract within this system is to check the IMEI and phone number associated with the GPS coordinates, and whether it is a registered user or not.

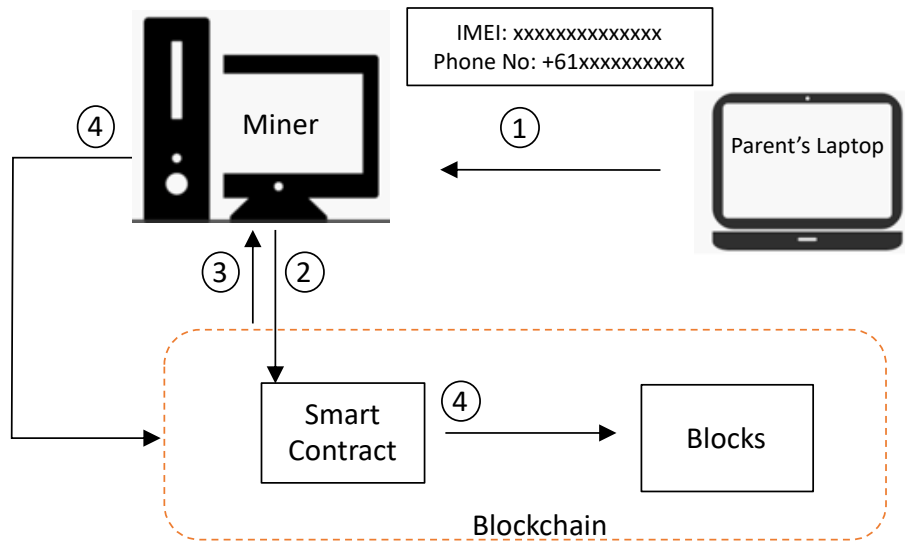

**Fig. 3.** Registering a device in the blockchain. This process follows four steps: 1. The IMEI and Phone number are sent to the blockchain via miner. 2. The miner checks the smart contract to verify the node can have access, if no then drop the request. 3. If yes then reply to the miner that the device can be added. 4. The miner adds the device's IMEI and Phone number to the blockchain.

### 4.3 Adding GPS coordinates to the blockchain

As shown in Fig 4 the RPi3 gets the GPS coordinates from the SIM900 module and then sends it to blockchain via miner node to add into the blockchain. The IMEI number and phone number are also sent to miner along with GPS coordinates against a timestamp. The miner verifies the node's address (RPi3), and registration of IMEI and phone number of the phone, and then adds them to blockchain.

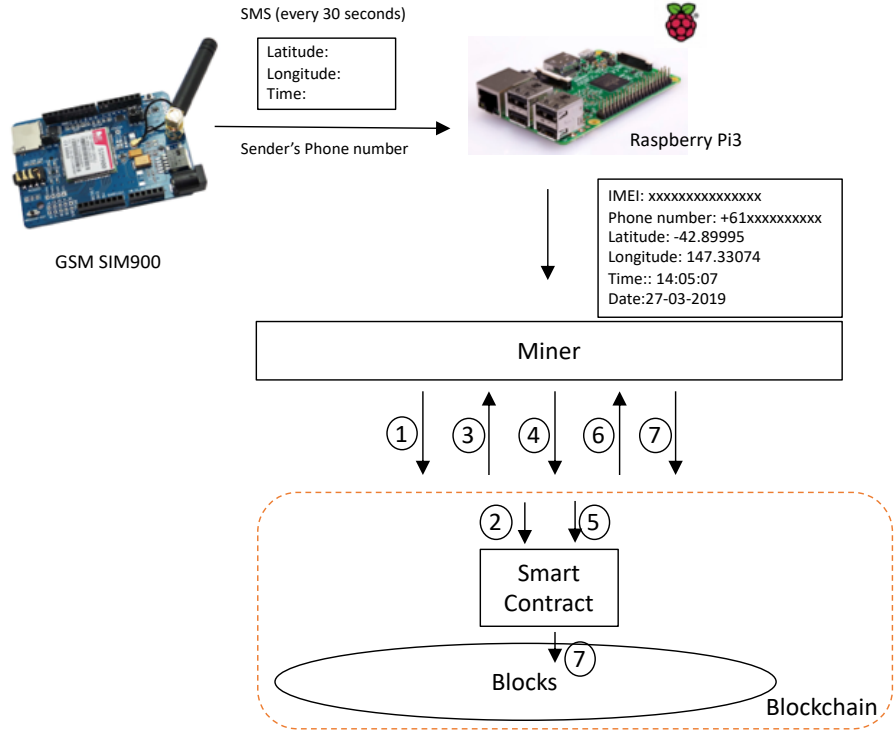

**Fig. 4.** Adding data in the blockchain. The process follows seven steps: 1. The miner requests whether the sender is a registered node. 2. The request is passed through a smart contract, if it returns negative it is dumped. 3. If it is a registered node then return true. 4. Check if the IMEI and phone number are registered. 5. The request is passed through the smart contract similar to step two. 6. If they are registered then return true. 7. GPS coordinates, node address, IMEI and phone number are added to the blockchain by miner via smart contract.

## 4.4 Reading GPS Coordinates from Blockchain

As shown in Fig 5 one of the features of a smart contract deployed in this private block-chain is to secure the access of blockchain. This blockchain only allows the parent's laptop (Node) to access/read the data from blockchain. Parents can also access all the relevant data (GPS coordinates against time) of a particular user by providing their IMEI and phone number.

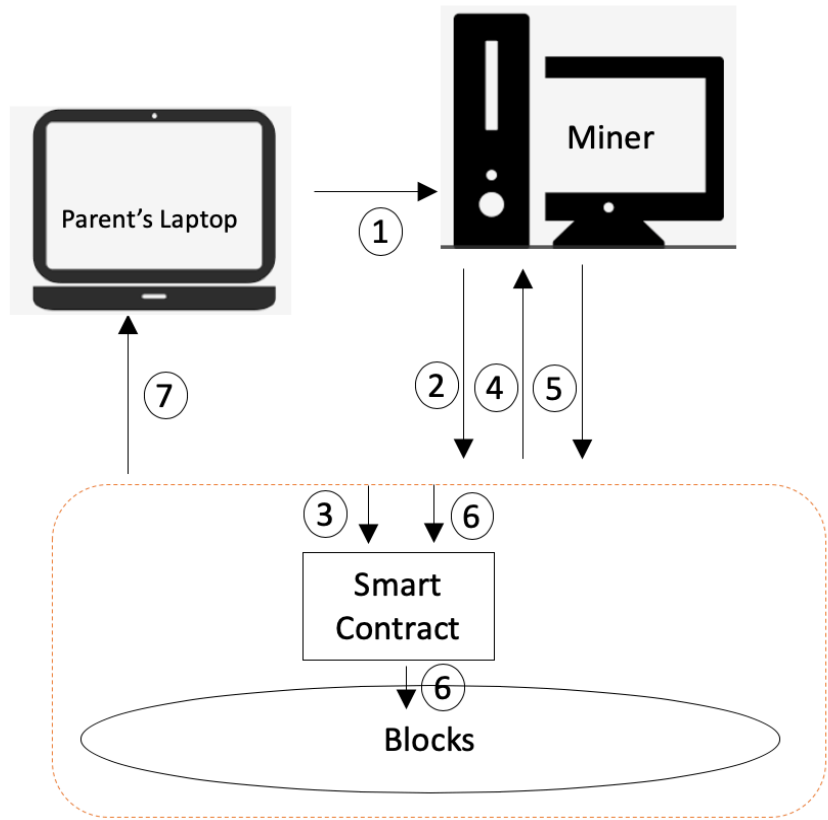

**Fig. 5.** Accessing data from the blockchain. This process follows seven steps: 1. A request is sent from the parent's node to the blockchain via miner seeking access to the data of a particular IMEI and phone number. 2. The miner checks the smart contract to check if the node has permission to access the data. 3. The smart contract executes, if no then rejects the request, if yes then returns true. 4. The miner acknowledges the true value for permission. 5. The miner sends the relevant related data to execute the search to the blockchain. 6. The smart contract executes the search of the blocks until the data is found. 7. Return the data to the parent's node.

## 5      Implementation and testing

### 5.1     Transmitter Side

The transmitter side consisted of an android phone with an app installed, the app was used to send the data from the phone to GSM SIM900 module. The SLA4 (Scripting Layer for Android) library was used, which gives an advantage of writing and executing scripts written in different scripting languages on Android. To run SLA4 scripts on android phones, QPython was installed on an android phone. Qpython is a script-running engine which is used to run SLA4 scripts on android. It also gives access to APIs in Android like SMS, GPS, etc.

Android uses a permission-based procedure which means that all the permissions an application needs are to be declared in a file named "AndroidManifest.xml". This lets the user of the phone know which exact permissions are needed by a specific application. In the case of this system, it needs to send and receive SMS messages and access GPS coordinates, so SMS and GPS related permissions were added in "AndroidManifest.xml" file:

```
<uses-permission android:name="android.permis-
sion.SEND_SMS"/>
  <uses-permission android:name="android.permis-
sion.RECEIVE_SMS"/>
  <uses-permission android:name="android.permis-
sion.ACESS_FINE_LOCATION"/>
```

The application needed to run on a background thread without affecting the UI lifecycles so an "IntentService" was created. This background service class allowed manipulation of long-running processes without affecting the responsiveness of the mobile phone's user interface. Python was used for scripting to send an SMS every 30 seconds, Fig 6 demonstrates an example of the content sent to the receiver (GSM SIM900).

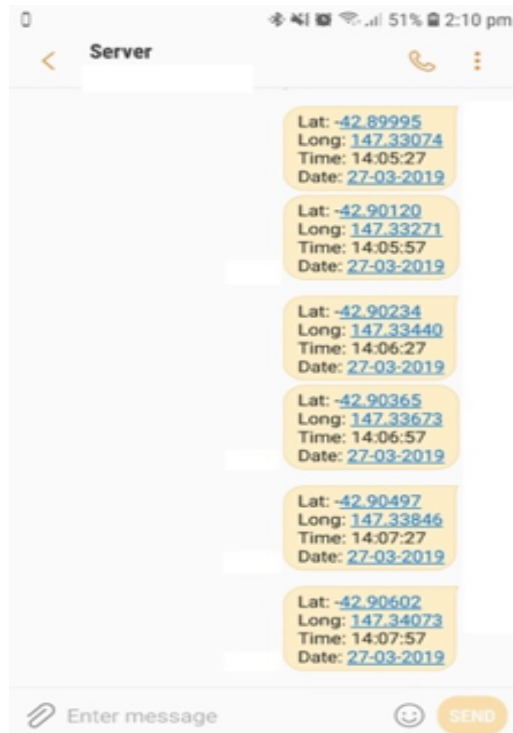

**Fig. 6.** The format of the text messages sent from the phone to the receiver. The format includes a latitude, longitude, timestamp as HH:MM:SS and date formatted as DD-MM-YYYY. The messages are sent in 30 second cycles. (Note: In this figure, server is the receiver i.e; RPi3)

### 5.2 Receiver Side

The receiver side consisted of the GSM SIM900 and an RPi3. GSM SIM900 is quad-band GSM/GPRS module that works on GSM 850MHz, EGSM 900MHz, DCS 1800MHz, and PCS 1900MHz [datasheet]. It uses a serial interface to interact with microcontrollers or other devices and accepts "AT" (Attention) commands, which instruct modem to execute its functions. A SIM card is required to attach to SIM900. The RPi3 was connected via GPIO (General Purpose Input/Output) pins to serial interface of SIM900. AT commands were used in the code to read incoming texts and send texts from SIM900.

To add GPS coordinates in to the blockchain, a private blockchain was set up with RPi3 as one node, while the parent's laptop and a server (miner) were two other nodes of the network. Blockchain was implemented on Ethereum using Geth (go-Ethereum), which is a command line interface to connect to Ethereum blockchain. Solidity was used as a compiler to compile smart contracts, and Truffle to test and deploy the smart contract on the blockchain.

Within our system the smart contract had the functions of:

- Checking the requests whether the particular requestee is authorized to make a particular request. For example, only the parent's node was allowed to register a new device in the blockchain and access all the relevant GPS coordinates against the timestamp of a particular handset.
- Another function of smart contract was to check whether the GPS coordinates are coming from a registered phone or not. If they were coming from a registered phone then they would be added to blockchain, if not then they would be discarded. RPi3 was only allowed to add data in to blockchain and had no access to read data from the blockchain.
- The third function of smart contract was to check whether the received coordinates are similar to that of the house or not. If they were similar, then a SMS would be sent to parent's phone to notify them that the particular family member had reached home. The inaccuracy of GPS receivers in mobile handsets had limited this project to not exactly compare the coordinates with that of the house but to use a range of a few meters from the house as the comparison.
- The fourth function was to provide GPS coordinates of a handset requested by the parent node by searching and filtering the data in transactions.

## 6    Limitations

This study created a closed system for an IoT environment that was designed to track via GPS a phone and relay that to a single board computer (RPi3) to be added in to a private blockchain. There were a number of limitations to this study, they involved the limits of the technology, the scope of testing, the integration into a mature IoT smart-home enabled system, and further application of artificial intelligence to overcome outages or topographical communication issues.

The limitations of the technology include the number of phones interfacing with the server for testing and verification of the system with two phones. It was assumed that a reasonable number of phones can be added to the system with no discernible delay, and future testing of the system will verify this. As the purpose was to test the validity of the proposed idea, the efficiency of the system wasn't a priority, as the project matures further improvements will be made to the technology.

The scope of testing involved a limited distance from a fixed-point using foot traffic as a mode of transport, this needs to be further tested in different modes of transport at different locations in the environment to check for flaws in the system, for example blackspots creating non-delivery messages.

At present this system was designed as part of a larger IoT enabled smart home system, but it was tested in a stand-alone environment, as the smart-home project matures the system will be integrated to leverage the blockchain with smart contracts and relevant components.

The application of artificial intelligence to the project wasn't the main goal, but the implementation of AI will be useful for predicting routes that could have been taken based on location data pulled from the blockchain. This may be required in instances of signal blackspots or other forms of network disruption. Furthermore, the use of smart contracts for using locational data may require a recommender system that can read from the chain and cross-check with familiar locations to identify anything notable. The use of artificial intelligence ties in to the larger smart-home project and its benefits are yet to be fully realized for blockchain based technology.

Finally, while this project has considered the implementation of technology to create immutable records, it hasn't as of yet considered the ethical implications of using blockchain for tracking. This will be the basis of future work relating to the smart-home IoT enabled project from the non-technical perspective of human computer interaction.

## 7 Conclusion and Future Work

In this paper a secure tracking system was proposed and implemented as a family security system based on blockchain technology. Where the primary goals were to have every aspect of the system controlled by the family and having the immutability of the data to prevent tampering at the easiest point of interference. This was done in a way that results in all relevant information about the location of family members remaining within the family smart-home system. This is in contrast to other tracking or security applications that offer similar services, where data is stored in some external server that could be accessed by entities other than family members. One possible extension of this project would involve adding artificial intelligence to the system and using location data to suggest nearby locations of relevance like restaurants, shops, or landmarks via SMS. Though the locational dataset for family members was too small, as such it wasn't feasible to train the system for this purpose. This system could be used in fleet monitoring and supply chain management to track vehicles. Future work will consider integrating this system with other smart home systems on blockchain in depth to come up with a smart home system based on blockchain.